\documentclass{article}

\usepackage{arxiv}

\usepackage[utf8]{inputenc} 
\usepackage[T1]{fontenc}    
\usepackage{hyperref}       
\usepackage{url}            
\usepackage{booktabs}       
\usepackage{amsfonts}       
\usepackage{nicefrac}       
\usepackage{microtype}      
\usepackage{lipsum}		
\usepackage{amsmath}
\usepackage{graphicx}
\usepackage{subfigure}
\usepackage{natbib}
\usepackage{doi}

\title{A new turbulence model based on scale decomposition}

\author{ \href{https://orcid.org/my-orcid?orcid=0000-0001-9711-1847}{\includegraphics[scale=0.06]{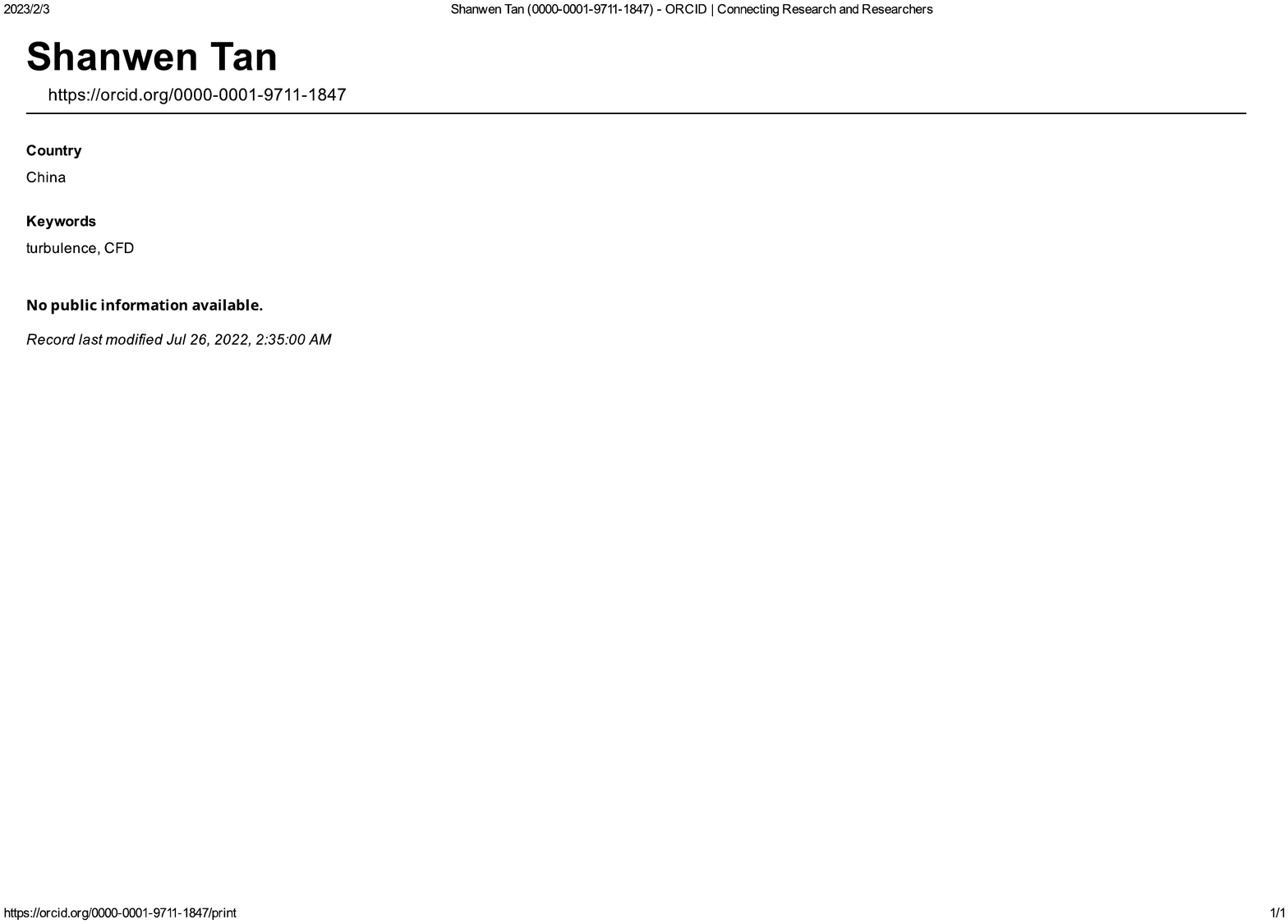}\hspace{1mm}Shanwen Tan}\thanks{ Key Laboratory of Fluid and Power Machinery, Ministry of Education, Xihua University, Chengdu  610039, China.} \\
	 School of Architecture and Civil Engineering\\
	Xihua University\\
	Chengdu Sichuan China, 610039 \\
	\texttt{tanshanwen68@mail.xhu.edu.cn} \\
}

\renewcommand{\shorttitle}{A new turbulence model...}

\hypersetup{
pdftitle={A new turbulence model based on scale decomposition},
pdfsubject={q-bio.NC, q-bio.QM},
pdfauthor={Shanwen Tan},
pdfkeywords={fluid mechanics, turbulence calculation, scale decomposition, flow around cylinder.},
}

\begin{document}
\maketitle

\begin{abstract}
 Based on the characteristics of the multi-scale and similarity at different scales in turbulent flow, we propose a scale decomposition for solving the turbulence problem of incompressible Newtonian fluid. The solution domain is decomposed into two-level scales, the large scale component represents mean flow and large scale eddies, and the small scale one represents the turbulent fluctuations. The problem is solved in large scale by the equations of motion and the effect of the turbulent fluctuations on the mean flow is evaluated approximately by using equivalent eddy. Furthermore, the effect of equivalent eddy is decomposed into two parts including convective effect and diffusion effect, which is expressed as a function of mean quantities in large scale. The modified Naiver-Stokes equations are established, there ensures the closure of the equations in large scale. Finally, the modified Naiver-Stokes equations is verified by the numerical simulation. Flow around cylinder is numerically investigated and able to obtain flow behavior from low to high Reynolds numbers. A general-purpose turbulence model is established in this study, which is worthy of engineering application.
\end{abstract}

\keywords{fluid mechanics \and turbulence calculation \and scale decomposition \and flow around cylinder}

\section{Introduction}
Turbulent flow exist widely in engineering fields and in nature. It is great challenge to obtain an accurate prediction of turbulent flow in simulation. The Naiver-Stokes equations (N-S equations) can accurately describe turbulent flows, the direct approach of solving the N-S equations is call direct numerical simulation (DNS)\cite{tennekes1972first}. In turbulent flow, the velocity, pressure and other fluid mechanical quantities  behave as dramatic fluctuations in space and time, and cover a wide range of length and time scales, these properties make the DNS of turbulent flows very difficult. Another alternative is to pursue a statistical approach, due to some statistic properties in turbulence, people are more concerned about averages of velocity, pressure and other quantities to describe the turbulent flow. At present, several different, useful averages play important roles in simulation. There are two categories of average operation in mainstream simulation\cite{versteeg2007introduction}.

One is Reynolds-averaged Navier-Stokes (RANS) equations. Before the application of numerical methods, the N-S equations are time averaged, due to the interactions between various turbulent fluctuations, extra terms appear in the time-averaged flow equations, and add additional turbulent shear stresses to the flow, which are known as the Reynolds stresses. In order to be able to compute turbulent flows with the RANS equations, these extra terms are modelled with appropriate turbulence models to close the RANS equations. There are a lot of turbulence models to be useful in a general-purpose computational fluid dynamics  (CFD) code. The computing resources required for reasonably accurate flow computations are modest, so this approach has been the mainstay of engineering flow calculations.

Large eddy simulation (LES) is another method based on average operation. Prior to the computations, this method use space filter to the N-S equations, the filter can pass the larger eddies and rejects the smaller eddies. The resolved flow include mean flow and large eddies. This is an intermediate form of turbulence calculations which tracks the behaviour of the larger eddies, the effects on the resolved flow due to smaller, unresolved eddies are included by means of a so-called sub-grid scale model. This technique is starting to address CFD problems with complex geometry. 

For RANS and LES, the closure of the equations depend on appropriate turbulence model. All the models were developed basing on semi-empirical and dimensional analysis method, the semi-empirical theories of turbulence are valuable for solving a number of important practical problems. However, the hypotheses adopted in these models have no reliable physical foundation and contribute little to the understanding of the physical nature of turbulence. Until now, there is no existence of universal
turbulence closure models \cite{davidson2015turbulence,adams2007mathematics}.

In this paper, we present a few ideas aiming at establishing a new turbulence model. It start from the multi-scale characteristics of turbulence and the invariance of scale transformation, before the computation, with a user-chosen scale as a interface, the solution space is decomposed into large scale subspace and small scale subspace, the large-scale components is solved directly, and the unsolved eddies in small scale was evaluated approximately by equivalent eddy, then some amendatory Navier-Stokes equations are establish in large scale sub-space, this ensures the closure of equations and realizes the calculation of the flow field. This method can obtain a general-purpose model of the turbulence flow.

\section{Model setup}
\label{sec:model}
Turbulence can be regarded as a spatially complex vorticity field which develops itself in a chaotic manner, fluid flow properties can be decompose into a steady mean value and some fluctuating components superimposed on mean flow. The mean flow acts as a mechanism for initiating the vorticity field, thereafter it played little role on the evolution of the vorticity field, however, there is a complex interaction between the vorticity and the mean flow. The mean flow generates, maintains and redistributes the vorticity, while the vorticity acts back on the mean flow, shaping the mean velocity redistribution. Turbulent flows fluctuate on a broad range of length and time scales. Based on Richardson's energy cascade, energy is passed from large to small scales by a repeated sequence of steps, large eddies obtain energy from the mean flow, and then break into smaller eddies through cascade process, this cascade is a multistage process, involving a hierarchy of eddies of varying size, and representing the multi-scale structure in the whole turbulent flow. In turbulent flow, the characteristics of large-scale eddies depend considerably on the geometry of the boundaries of flow and hence will be very different for different types of flow, the characteristics of small-scale eddies, in general, have less connection to the boundaries of flow and possess some universal characters\cite{frisch1995turbulence,she1994universal}.

The large-scale eddies and the mean flow combine into the large-scale components which make key contribution to the transfer of momentum and heat in turbulent flow. Therefore, it is natural that in the development of the calculation method of turbulence attention should be first given to an accurate prediction of the large-scale components\cite{barenblatt1996scaling,barenblatt1998new,monin2013statistical}.

Based on this idea, we propose a scale decomposition. first, one scale is chosen properly, the solution space is decomposed into large scale sub-space and small scale sub-space based on user-chosen scale. 
In general, the user-chosen scale is the integral scale of the turbulence \cite{frisch1995turbulence,pope2000turbulent}, this can capture accurately large-scale components, meanwhile, ensure that the small-scale eddies have universal characters. In large scale sub-space, we seek to predict accurately both the mean flow and the evolution of all of the large-scale eddies using governing equations. The unresolved eddies in small-scale sub-space are parameterized using mean quantities in large scale, which ensure the closure of the governing equations.

First, the equations of fluid motion is established in large scale. According to the form invariance of momentum equations under scale transformation, the equations of fluid motion in large scale are formally the Navier-Stokes equations, only the relevant physical quantities are defined in large scale\cite{she1998universal,chorin1994vorticity,barenblatt1993scaling}. In addition, the effect of eddies in small scale must be considered. 

\begin{figure}[htbp]
\centering
\includegraphics[width=0.6\textwidth]{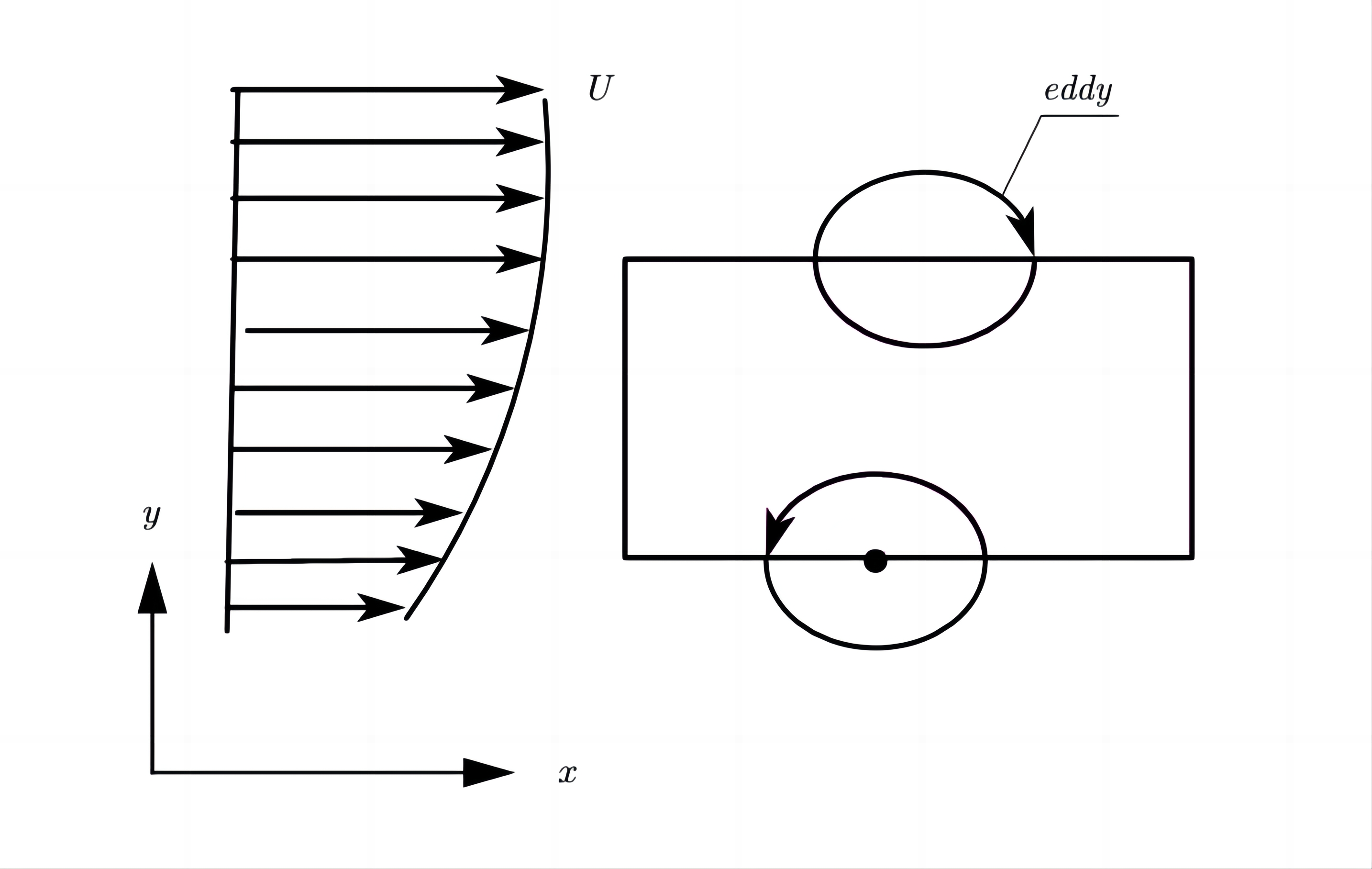}
\caption{Effect of eddy on flow field in two-dimensional control volume.}
\label{fig1} 
\end{figure}

How eddies in small scale affect mean flow. Refer to Figure \ref{fig1}, we consider a control volume in a two-dimensional turbulent shear flow parallel to the x-axis with a mean velocity gradient in the y-direction. The presence of vortical eddy motions creates strong mixing, these motions cannot create or destroy mass, but fluid parcels transported by the eddies will carry momentum and energy into and out of the control volume, because of the existence of the velocity gradient, the eddy passing through the interface causes additional momentum exchange in the control volume, which causes the faster moving fluid layers to be decelerated and the slower moving layers to be accelerated. Consequently, it changes the velocity distribution of the local flow field. the fluid layers experience additional turbulent shear stresses, which are known as the Reynolds stresses\cite{versteeg2007introduction}.

This suggests that the equations for momentum should be affected by the appearance of eddy. we do not track the effect of different eddy alone, but consider the cumulative effect of all eddies in small scale.

The cumulative effect is defined as function $R\left( \boldsymbol{U} \right)$, this function is added to the Navier-Stokes equations in large scale. We restrict to incompressible, viscous and Newtonian fluid, and the governing equations in large scale are

\begin{eqnarray} \label{eq1}
\nabla \left( \boldsymbol{U} \right) =0
\end{eqnarray}

\begin{eqnarray} \label{eq2}
  \rho \left( \frac{\partial \boldsymbol{U}}{\partial T}+\left( \boldsymbol{U}\cdot \nabla \right) \boldsymbol{U} \right) =-\nabla \boldsymbol{P}+\rho \boldsymbol{G}+\mu \nabla ^2\boldsymbol{U}+R\left( \boldsymbol{U} \right)  
\end{eqnarray}
where $T$ is the time, $\boldsymbol{U}$ and $\boldsymbol{P}$  are the velocity and pressure of the flow field in large scale,  and $\mu$ is dynamic viscosity of the fluid.

Here, take the momentum equation of X-direction as an example to show how to establish a complete governing equation in large scale.

\begin{eqnarray} \label{eq3}
  \rho \left( \frac{\partial U}{\partial T}+U\frac{\partial U}{\partial X}+V\frac{\partial U}{\partial Y}+W\frac{\partial U}{\partial Z} \right) =-\frac{\partial P}{\partial X}+\rho G_X+
\mu \left( \frac{\partial ^2U}{\partial X^2}+\frac{\partial ^2U}{\partial Y^2}+\frac{\partial ^2U}{\partial Z^2} \right) +R\left( U \right)  
\end{eqnarray}
where ${R\left( U \right)}$ is the component in the X-direction of the $R\left( \boldsymbol{U} \right)$. 

To ensure closure the governing equation, ${R\left( U \right)}$ must be expressed as a function of the mean quantities in large scale. In fluid flow, there are two basic processes: convection and diffusion. The cumulative effect of eddies on mean flow can only be  exerted  by the convection and diffusion, so ${R\left( U \right)}$ can be decomposed into the effect of the convection and of diffusion, as shown below

\begin{eqnarray} \label{eq4}
R\left( U \right) =R_1\left( U \right) +R_2\left( U \right)
\end{eqnarray}

Substitute Equation (\ref{eq4}) into equation (\ref{eq3}) to get equation

\begin{eqnarray} \label{eq5}
\rho \left( \frac{\partial U}{\partial T}+U\frac{\partial U}{\partial X}+V\frac{\partial U}{\partial Y}+W\frac{\partial U}{\partial Z} \right) +R_1\left( U \right) =-\frac{\partial P}{\partial X}+\rho G_X \nonumber
\\
+\mu \left( \frac{\partial ^2U}{\partial X^2}+\frac{\partial ^2U}{\partial Y^2}+\frac{\partial ^2U}{\partial Z^2} \right) +R_2\left( U \right)
\end{eqnarray}

Now, we ascertain function ${R_1\left( U \right)}$.
\begin{figure}[htbp]
\centering
\includegraphics[width=0.6\textwidth]{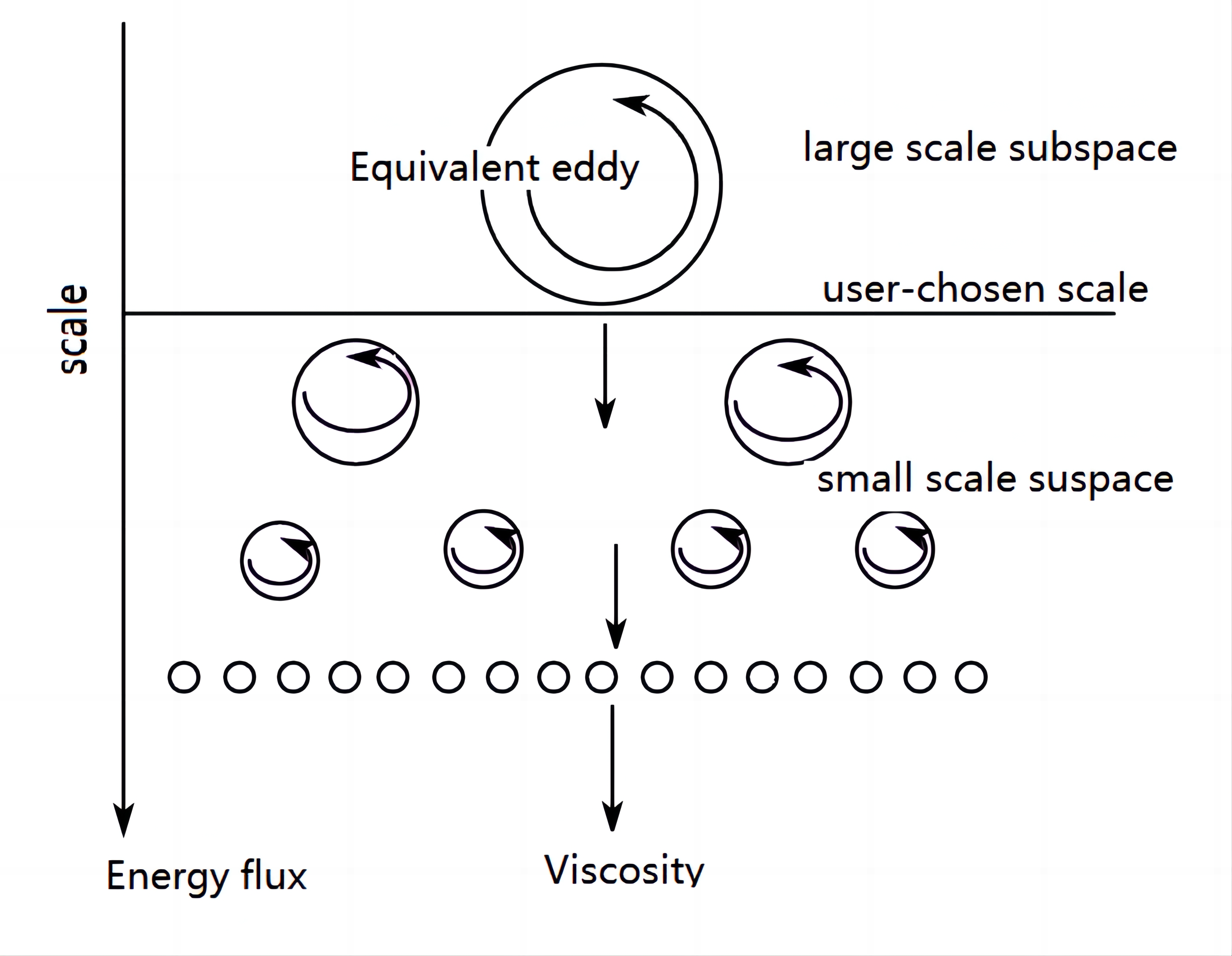}
\caption{  A schematic representation of the energy cascade.}
\label{fig2}
\end{figure}

Eddy changes the momentum and energy distribution of the local flow field with the aid of the convection of eddy. It is known that there is minimal-sized distinguishable eddy at every scales. The minimal-sized distinguishable eddy in large scale is shown in figure \ref{fig2}, eddies in small scale are produced by the cascade process from this eddy, therefore, according to the energy transfer path of each eddy, all the energy of eddies in small scale come from this minimal-sized distinguishable eddy. we can use this eddy to evaluate the cumulative effect of eddies in small scale, we call this minimal-sized distinguishable eddy as equivalent eddy\cite{gioia2017spectral,gioia2010spectral}.

 Then we determine the size of the equivalent eddy. A series of eddies in turbulent flow are formed by cascade process, which is not endless. There exists the smallest scale of motion in a turbulent flow, the energy associated with eddy motions less than the smallest scale is dissipated and converted into thermal internal energy. The eddy size of the smallest scale is proportional to $\nu ^{3/4}$ \cite{griebel1998numerical,rota1976encyclopedia}. The eddy size of the smallest scale is only related to kinematic viscosity of the fluid and independent of the position of the eddy .

 For standard Navier-Stokes equations, there exhibits a property known as scale in-variance. Suppose $u\left( x,t \right)$ represents one solution of the Navier-Stokes equations. Then $u'\left( x',t' \right)$ is also a solution provided that
 
 \begin{eqnarray} \label{eq6}
	t'\rightarrow \lambda ^{1-H}t,\ \ r'\rightarrow \lambda r,\ \ u'\rightarrow \lambda ^Hu,\ \ \nu '\rightarrow \lambda ^{1+H}\nu
\end{eqnarray}

 where $\lambda$ is scale factor, reflecting the amplification or reduction of the scale, $H$ is a scaling exponent . For the fully developed turbulent in inertia region $H=1/3$ .

  Thus a solution at one scale has its counterparts at all other scales. This has led to some speculation that the equivalent kinematic viscosity of the fluid is $\lambda ^{1+H}\nu$ in large scale, the equivalent eddy is the smallest eddy in large scale and its size can be expressed proportional to $\left( \lambda ^{1+H}\nu \right) ^{3/4}$ \cite{davidson2015turbulence}. The size of equivalent eddy is only related to kinematic viscosity and scale, and independent of the position of the eddy. 

In numerical simulations, it is assumed that the characteristic dimension of mesh is $\Delta l$ , then the diameter of the equivalent eddy is proportional to $\Delta l$, it can be express as $d=k\left(\nu \right)\Delta l$ , The value of $k\left(\nu \right)$ can be determined by numerical experiences and experimental measurement.

In this new model, we assume that the calculation point is accompanied by an equivalent eddy with the center as the calculation point. This eddy will produce extra momentum exchange at the calculation point, the corresponding convection items in equation need to be corrected, then the equation (\ref{eq5}) can be corrected to equation as

\begin{eqnarray} \label{eq7}
\rho \left( \frac{\partial U}{\partial T}+U\left( \frac{\partial U}{\partial X}+R_{11}\left( U \right) \right) +V\left( \frac{\partial U}{\partial Y}+R_{12}\left( U \right) \right) +W\left( \frac{\partial U}{\partial Z}+R_{13}\left( U \right) \right) \right) =\nonumber
\\
-\frac{\partial P}{\partial X}+\rho G_X+\mu \left( \frac{\partial ^2U}{\partial X^2}+\frac{\partial ^2U}{\partial Y^2}+\frac{\partial ^2U}{\partial Z^2} \right) +R_2\left( U \right)
\end{eqnarray}

Now using $R_{12}\left( U \right)$ as an example determines the specific expression.

\begin{figure}
\centering
\includegraphics[width=0.6\textwidth]{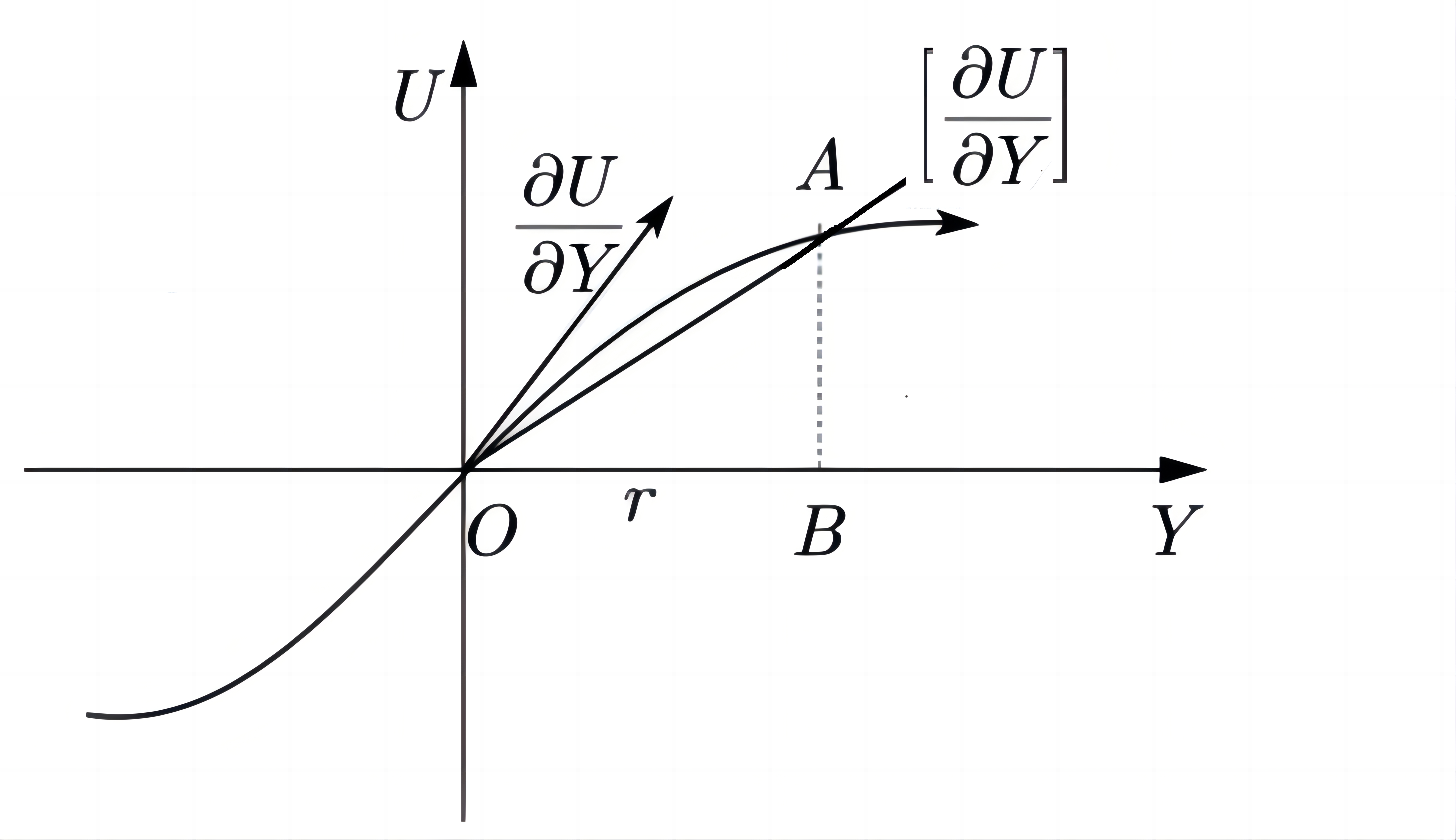}
\caption  {the mean velocity in y-direction and its redistribution due to motion of equivalent eddy.}
\label{fig3}
\end{figure}

Figure \ref{fig3} shows the distribution of the mean velocity in y-direction and its change due to  rotation of the eddy. Where, $O$ is the center of the equivalent eddy,  $r$ is the radius of the eddy, the curve is the distribution of the mean velocity, and $A$ is separate point of the interface between the eddy and the mean flow field. With the center of the eddy as a calculation point, the rotation of the eddy does not destroy the continuity of velocity, but the velocity distribution is changed near the center of the eddy , the curve $\widetilde{OA}$ , the original mean velocity distribution , has changed to $\overline{OA}$ at scope of the eddy. 

Assuming that the velocity of the eddy center is $ U \left(O \right)$ , then the velocity of point $A$ can be expressed as $U \left(r \right)$ . Using the Taylor series,  $U \left(r\right)$ can be expressed as equation below,  here only the first three items are retained.

\begin{eqnarray} \label{eq8}
U\left( r \right) =U\left( O \right) +\frac{\partial U}{\partial Y}r+\frac{1}{2!}\frac{\partial ^2U}{\partial Y^2}r^2+...
\end{eqnarray}

Due to rotation of the eddy, the velocity distribution near the center of the eddy change from $\widetilde{OA}$ to $\overline{OA}$ at scope of the eddy. Its actual first derivative of the velocity is

\begin{eqnarray} \label{eq9}
    \left[ \frac{\partial U}{\partial Y} \right] =\frac{U\left( r \right) -U\left( O \right)}{r}=\frac{\partial U}{\partial Y}+\frac{1}{2!}\frac{\partial ^2U}{\partial Y^2}r
\end{eqnarray}

The coefficient of the second derivative term is defined as a new parameter $\delta$ , which $\delta=\frac{1}{2!} r$, we call it as scale coefficient, and equation (\ref{eq9}) is rewritten as

\begin{eqnarray} \label{eq10}
\left[ \frac{\partial U}{\partial Y} \right] =\frac{\partial U}{\partial Y}+\delta \frac{\partial ^2U}{\partial Y^2}
\end{eqnarray}

Therefore, the correction term can be obtained as
\begin{eqnarray}\label{eq11}
R_{12}\left( U \right) =\delta \frac{\partial ^2U}{\partial Y^2}
\end{eqnarray}

Similarly,
\begin{eqnarray} \label{eq12}
R_{13}\left( U \right) =\delta \frac{\partial ^2U}{\partial Z^2}
\end{eqnarray}

Since the rotational motion of the eddy do not change the radial velocity distribution of the eddy. So
\begin{eqnarray} \label{eq13}
R_{11}\left( U \right) =0
\end{eqnarray}

Substituting equations (\ref{eq11}), (\ref{eq12}) and (\ref{eq13}) into equation (\ref{eq7}), we can get

\begin{eqnarray} \label{eq14}
\rho \left( \frac{\partial U}{\partial T}+U\left( \frac{\partial U}{\partial X} \right) +V\left( \frac{\partial U}{\partial Y}+\delta \frac{\partial ^2U}{\partial Y^2} \right) +W\left( \frac{\partial U}{\partial Z}+\delta \frac{\partial ^2U}{\partial Z^2} \right) \right) =\nonumber \\
-\frac{\partial P}{\partial X}+\rho G_X
+\mu \left( \frac{\partial ^2U}{\partial X^2}+\frac{\partial ^2U}{\partial Y^2}+\frac{\partial ^2U}{\partial Z^2} \right) +R_2\left( U \right)
\end{eqnarray}

Below we determine the function $R_2 \left(U \right)$ .

To facilitate the derivation of the formula, the viscous terms of the governing equation are expressed in the form of stress. The momentum equation is as follows

\begin{eqnarray} \label{eq15}
\rho \left( \frac{\partial U}{\partial T}+U\frac{\partial U}{\partial X}+V \frac{\partial U}{\partial Y} +W\frac{\partial U}{\partial Z}  \right) =-\frac{\partial P}{\partial X}+\rho G_X
+\frac{\partial \tau _{XX}}{\partial X}+\frac{\partial \tau _{YX}}{\partial Y}+\frac{\partial \tau _{ZX}}{\partial Z}
\end{eqnarray}

For Newtonian fluid, the constitutive equations are

\begin{eqnarray} \label{eq16}
\tau _{XX}=2\mu \frac{\partial U}{\partial X}\ \ \ \ \tau _{YX}=\mu \left( \frac{\partial U}{\partial Y}+\frac{\partial V}{\partial X} \right) \ \ \ \ \tau _{ZX}=\mu \left( \frac{\partial U}{\partial Z}+\frac{\partial W}{\partial X} \right)
\end{eqnarray}

Substitute equation (\ref{eq16}) into equation (\ref{eq15}) to get the following equation

\begin{eqnarray} \label{eq17}
\rho \left( \frac{\partial U}{\partial T}+U\frac{\partial U}{\partial X}+V\frac{\partial U}{\partial Y}+W\frac{\partial U}{\partial Z} \right) =-\frac{\partial P}{\partial X}+\rho G_X
+\frac{\partial}{\partial X}\left( 2\mu \frac{\partial U}{\partial X} \right) +\nonumber\\
\frac{\partial}{\partial Y}\left( \mu \left( \frac{\partial U}{\partial Y}+\frac{\partial V}{\partial X} \right) \right) +\frac{\partial}{\partial Z}\left( \mu \left( \frac{\partial U}{\partial Z}+\frac{\partial W}{\partial X} \right) \right)
\end{eqnarray}

The viscous terms contain the first partial derivative of the velocity, these first derivative terms need to be replaced to be actual first derivatives as list at equation (\ref{eq10}) and some similar expressions, then

\begin{eqnarray} \label{eq18}
\rho \left( \frac{\partial U}{\partial T}+U\frac{\partial U}{\partial X}+V\left( \frac{\partial U}{\partial Y}+\delta \frac{\partial ^2U}{\partial Y^2} \right) +W\left( \frac{\partial U}{\partial Z}+\delta \frac{\partial ^2U}{\partial Z^2} \right) \right) =
-\frac{\partial P}{\partial X}+\rho G_X+\frac{\partial}{\partial X}\left( 2\mu \frac{\partial U}{\partial X} \right)+\nonumber \\
\frac{\partial}{\partial Y}\left( \mu \left( \left( \frac{\partial U}{\partial Y}+\delta \frac{\partial ^2U}{\partial Y^2} \right) +\left( \frac{\partial V}{\partial X}+\delta \frac{\partial ^2V}{\partial X^2} \right) \right) \right) +\nonumber \\
\frac{\partial}{\partial Z}\left( \mu \left( \left( \frac{\partial U}{\partial Z}+\delta \frac{\partial ^2U}{\partial Z^2} \right) +\left( \frac{\partial W}{\partial X}+\delta \frac{\partial ^2W}{\partial X^2} \right) \right) \right)
\end{eqnarray}

The continuity equation is

\begin{eqnarray} \label{eq19}
\frac{\partial U}{\partial X}+\frac{\partial V}{\partial Y}+\frac{\partial W}{\partial Z}=0
\end{eqnarray}

By substituting the continuity equation (\ref{eq19}) into equation (\ref{eq18}), the governing equation is finally obtained through appropriate simplification and re-arrangement.

\begin{eqnarray} \label{eq20}
\rho \left( \frac{\partial U}{\partial T}+U\frac{\partial U}{\partial X}+V\frac{\partial U}{\partial Y}+W\frac{\partial U}{\partial Z} \right) =-\frac{\partial P}{\partial X}+\rho G_X+\mu \left( \frac{\partial ^2U}{\partial X^2}+\frac{\partial ^2U}{\partial Y^2}+\frac{\partial ^2U}{\partial Z^2} \right)\nonumber\\
+{\delta \rho \left( V\frac{\partial ^2U}{\partial Y^2}+W\frac{\partial ^2U}{\partial Z^2} \right) }+{\delta \mu \left( -\frac{\partial ^3U}{\partial X^3}+\frac{\partial ^3U}{\partial Y^3}+\frac{\partial ^3U}{\partial Z^3} \right) }
\end{eqnarray}
where
\begin{eqnarray*}
R_1\left(U\right)=\delta \rho \left( V\frac{\partial ^2U}{\partial Y^2}+W\frac{\partial ^2U}{\partial Z^2} \right) \\
R_2\left(U\right)=\delta \mu \left( -\frac{\partial ^3U}{\partial X^3}+\frac{\partial ^3U}{\partial Y^3}+\frac{\partial ^3U}{\partial Z^3} \right)
\end{eqnarray*}

In these equations, $R_1 \left (U \right)$ represents the effect of fluctuating components on the mean quantities through convection, and  $R_2 \left (U \right)$ represents the effect of fluctuating components on the mean quantity through diffusion, these two terms together represent the cumulative effect of fluctuating components on the mean quantities.

Similar governing equations can be obtained in the Y-axis direction and Z-axis direction.

The equations of motion are rewritten in nondimensional form and the following formulas are obtained

(1) continuity equation

\begin{eqnarray} \label{eq21}
\frac{\partial U}{\partial X}+\frac{\partial V}{\partial Y}+\frac{\partial W}{\partial Z}=0
\end{eqnarray}

(2) momentum equations

X-axis direction
\begin{eqnarray} \label{eq22}
\frac{\partial U}{\partial T}+U\frac{\partial U}{\partial X}+V\frac{\partial U}{\partial Y}+W\frac{\partial U}{\partial Z}=-\frac{\partial P}{\partial X}+G_X+\frac{1}{Re}\left( \frac{\partial ^2U}{\partial X^2}+\frac{\partial ^2U}{\partial Y^2}+\frac{\partial ^2U}{\partial Z^2} \right)\nonumber\\
+\frac{\delta}{L}\left( V\frac{\partial ^2U}{\partial Y^2}+W\frac{\partial ^2U}{\partial Z^2} \right) +\frac{\delta}{L}\frac{1}{Re}\left( -\frac{\partial ^3U}{\partial X^3}+\frac{\partial ^3U}{\partial Y^3}+\frac{\partial ^3U}{\partial Z^3} \right)
\end{eqnarray}

Y-axis direction
\begin{eqnarray} \label{eq23}
\frac{\partial V}{\partial T}+U\frac{\partial V}{\partial X}+V\frac{\partial V}{\partial Y}+W\frac{\partial V}{\partial Z}=-\frac{\partial P}{\partial Y}+G_Y+\frac{1}{Re}\left( \frac{\partial ^2V}{\partial X^2}+\frac{\partial ^2V}{\partial Y^2}+\frac{\partial ^2V}{\partial Z^2} \right)\nonumber\\
+\frac{\delta}{L}\left( U\frac{\partial ^2V}{\partial X^2}+W\frac{\partial ^2V}{\partial Z^2} \right) +\frac{\delta}{L}\frac{1}{Re}\left( \frac{\partial ^3V}{\partial X^3}-\frac{\partial ^3V}{\partial Y^3}+\frac{\partial ^3V}{\partial Z^3} \right)
\end{eqnarray}

Z-axis direction
\begin{eqnarray} \label{eq24}
\frac{\partial W}{\partial T}+U\frac{\partial W}{\partial X}+V\frac{\partial W}{\partial Y}+W\frac{\partial W}{\partial Z}=-\frac{\partial P}{\partial Z}+G_Z+\frac{1}{Re}\left( \frac{\partial ^2W}{\partial X^2}+\frac{\partial ^2W}{\partial Y^2}+\frac{\partial ^2W}{\partial Z^2} \right)\nonumber\\
+\frac{\delta}{L}\left( U\frac{\partial ^2W}{\partial X^2}+V\frac{\partial ^2W}{\partial Y^2} \right) +\frac{\delta}{L}\frac{1}{Re}\left( \frac{\partial ^3W}{\partial X^3}+\frac{\partial ^3W}{\partial Y^3}-\frac{\partial ^3W}{\partial Z^3} \right)
\end{eqnarray}
Where $\delta$ is the scale coefficient, $L$ is the characteristic dimension of solution domain, and $Re$ is the Reynolds number.

Equations (\ref{eq21})-(\ref{eq24}) constitute the modified N-S equations at the large scale, and can be solved with appropriate initial and boundary conditions. The extra terms $R\left(\boldsymbol{U}\right)$ at the equation can be understood as Reynolds stress, and show anisotropy characteristics of turbulence flow. The governing equations built on scale decomposition ensure that the closure of equation and can be used to solve the turbulent flow problems of incompressible Newtonian fluids.

\section{Numerical Simulation}
\label{sec:numerical}
Flow around cylinders has been the topic of numerous experimental and numerical investigations because of its significance in engineering projects. It is geometrically very simple but exhibits numerous important physical phenomena, such as flow separation, vortex shedding, and turbulence. It is a good example to verify CFD \cite{yuce2016numerical,lakehal1997calculation}. 

Here simulation is restrict to the two-dimensional domain, and discretized by equidistant difference grids, the unknown variables are located at staggered grids. The convective terms adopt the upwind scheme and the viscous terms adopt the central difference scheme, the time term is discretized in Euler's first-order scheme. Using simple extension to approximately treat flows in arbitrary two-dimensional geometries, the cells of solution domain are divided into fluid cells  and obstacle cells, the modified N-S equations are then solved only in the fluid cells by the projection method in Chorin form. This algorithm implemented in detail is refer to \cite{griebel1998numerical}.

\begin{figure}[htbp]
\centering
\includegraphics[width=0.6\textwidth]{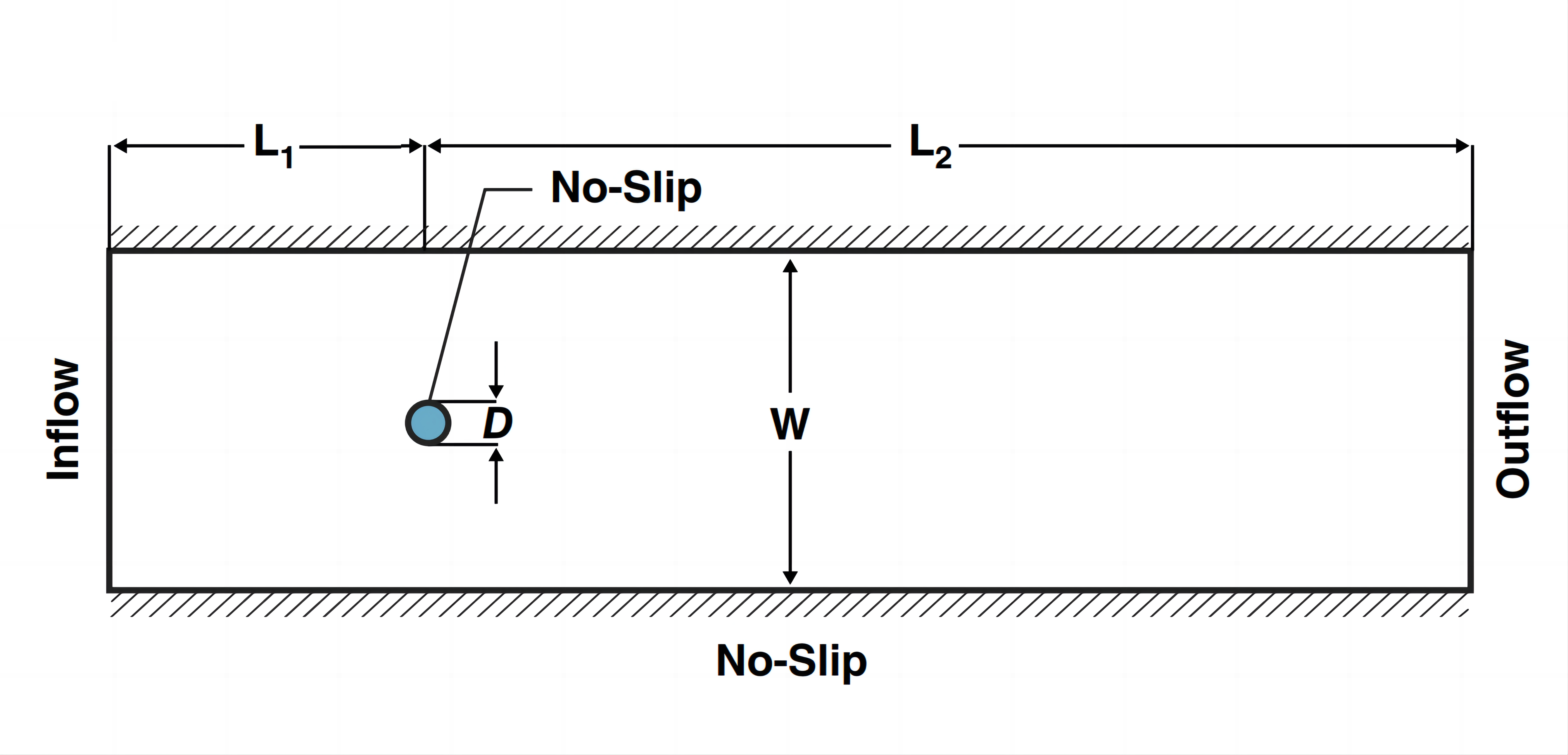}
\caption{ Geometric dimensions and boundary conditions of computational model.}
\label{fig4}
\end{figure}

The computational domain and the boundary conditions are shown in figure \ref{fig4}. The diameter of circular cylinders is characteristic length, let D=1m. The length of the channel is 40D with L1=4D, L2=36D,  and the height is 8D, the cylinder is immersed in the center of the channel.

The computational domain is discretized using equidistant grids, with the number of grids in the length direction being 800 and the height direction being 160. The inside of the cylinder is approximated by marking as obstacle cell. 

the governing equations after discretization is solved by the projection method, the velocity, pressure and other flow properties of the whole computational domain can be obtained. 

Before solving, the present algorithm needs to determine the value of the scale coefficient $\delta$ .  The scale coefficient is associated with the user-chosen scale and the kinematic viscosity of the fluid. According to the previous discussion, the characteristic dimension of the mesh is taken as the minimum value of the grid in two directions, and the scale coefficient $\delta$ is obtained by

\begin{eqnarray*}
\delta =k\cdot \min \left( \varDelta x,\ \varDelta y \right) /4
\end{eqnarray*}

Based on the existing numerical experience, In general, $k$ is between $1 \sim 3$, The value is also increased appropriately with the increase of $\text{Re}$.

In general, the streamline of the flow around a cylinder is only depend on $\text{Re}$ . At low  $\text{Re}$ , the inertial force is secondary to the viscous force in the flow field, and the streamline in both upstream and downstream of the cylinder remains symmetrical. With the increase of  $\text{Re}$, the inertia force slowly increases to the dominant position, and the streamlines in upstream and downstream of the cylinder gradually lose symmetry. 

\begin{figure}[htbp]
\centering
\includegraphics[width=0.8\textwidth]{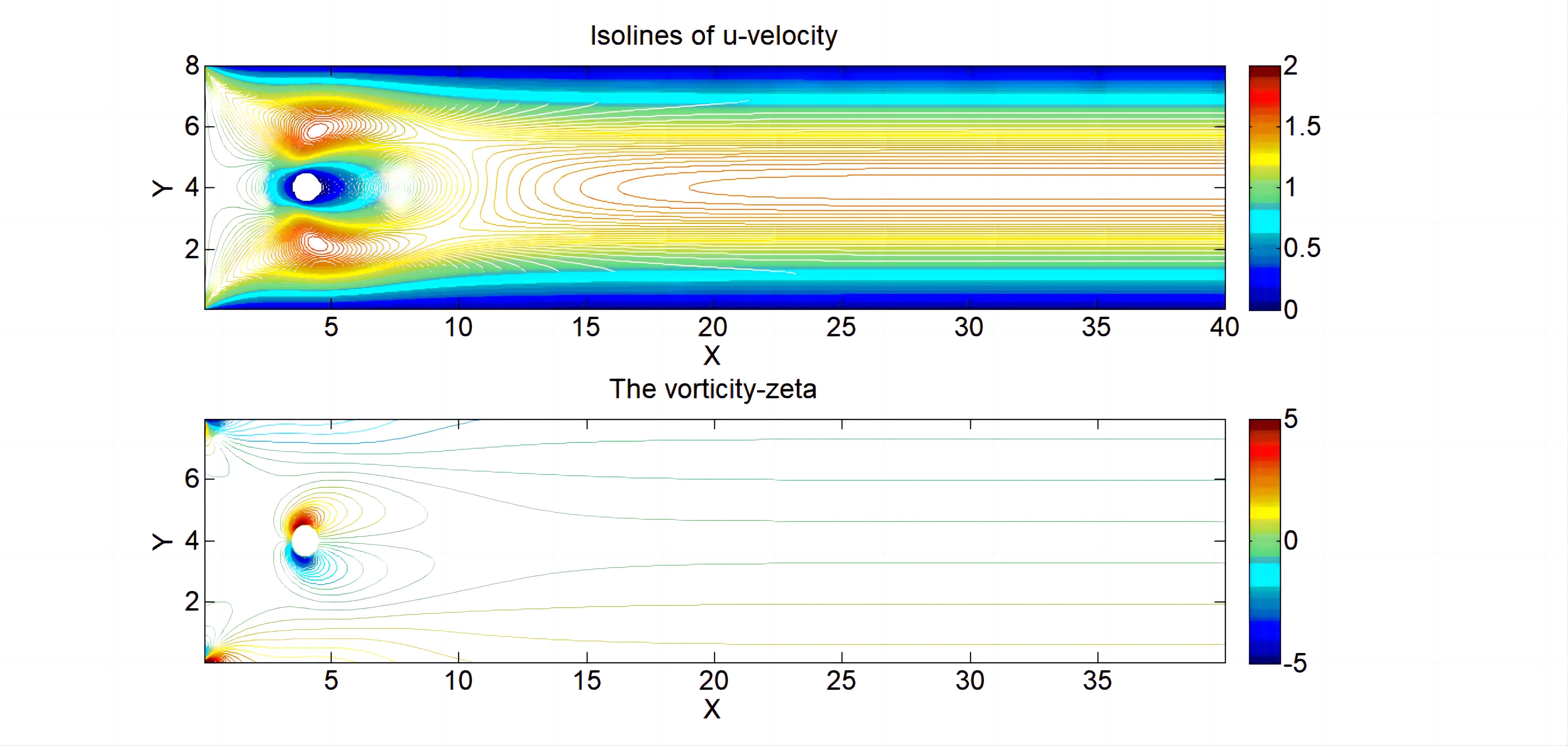}
\caption{   Contour lines of velocity distribution and vorticity distribution at Re=4.}
\label{fig-re4}
\end{figure}

Figure \ref{fig-re4} shows the velocity and vorticity distribution with  $\text{Re=4}$ . It can be seen that the velocity distribution continues to be symmetrical up and down.

As the increase of  $\text{Re}$, the fluid flowing along the surface of the cylinder begins to separate near the upper and lower vertices of the cylinder, forming a pair of fixed and symmetrical vortices downstream of the cylinder.

\begin{figure}[htbp]
\centering
\includegraphics[width=0.8\textwidth]{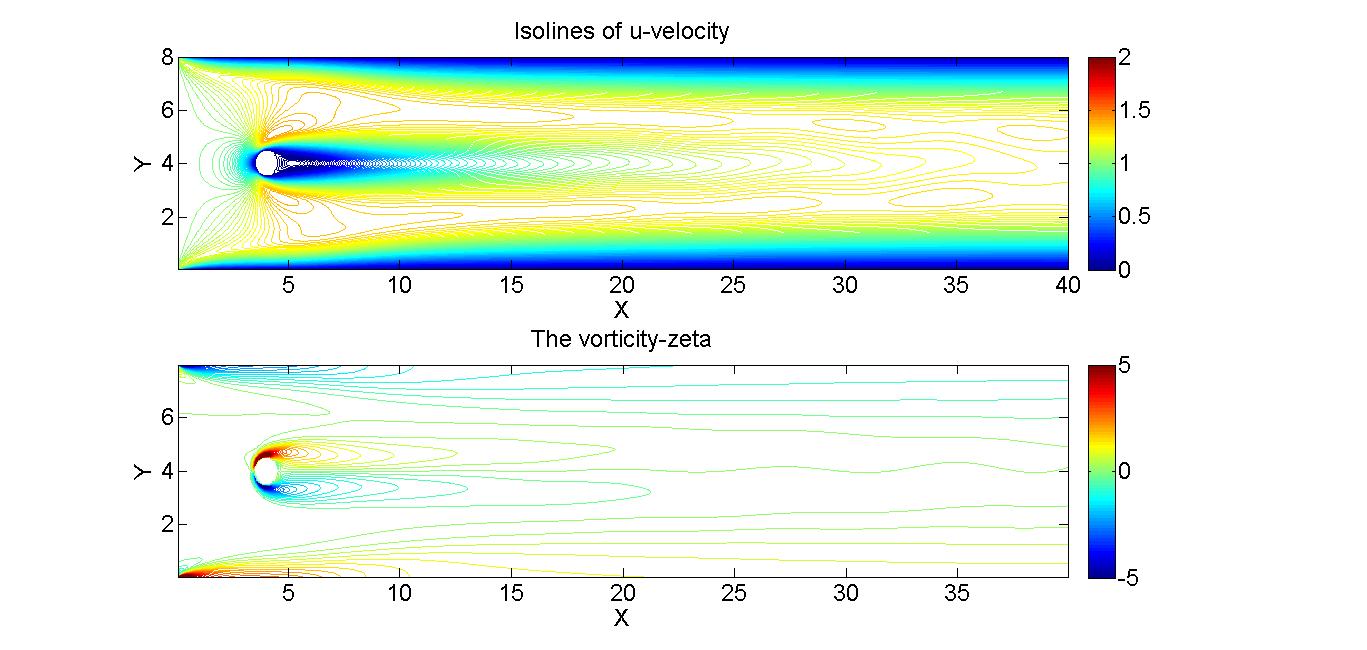}
\caption{    Contour lines of velocity distribution and vorticity distribution at Re=40.}
\label{fig-re40}
\end{figure}

Figure \ref{fig-re40} shows the velocity and vorticity distribution with  $\text{Re=40}$. It can be seen from the vorticity distribution diagram, there is a pair of symmetrical and opposite direction vortices behind the upper and lower vertices of the cylinder, and in the downstream region, the flow begins to lose symmetrical and form the wake zone.

\begin{figure}[htbp] 
\centering \includegraphics[width=0.8\textwidth]{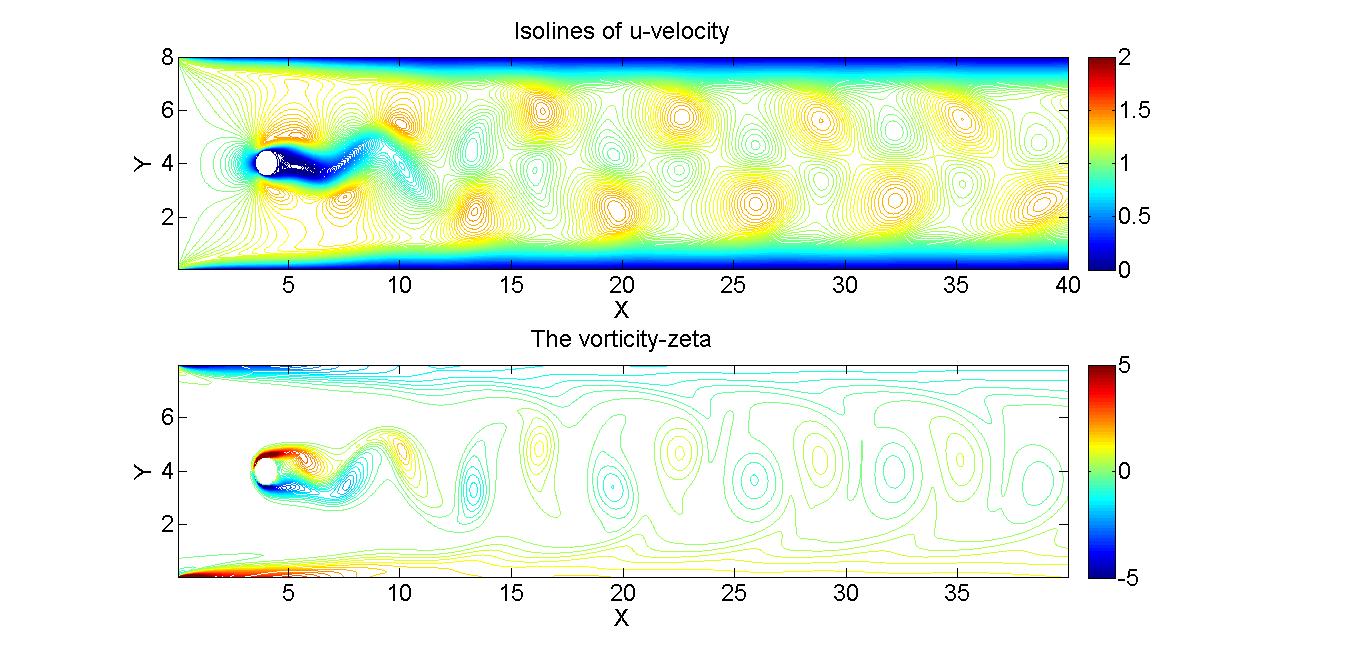} 
\caption{ Contour lines of velocity distribution and vorticity distribution at Re=100.}
\label{fig-re100} 
\end{figure}

When  $\text{Re}$ exceeds one critical value, the symmetrical vortex attached to the cylinder begins to lose stability, and the upper and lower sides of the rear edge of the cylinder periodically peel off, forming a regular vortex array, which is called Karman vortex street. Figure \ref{fig-re100} shows obvious characteristics of the vortex street.

\begin{figure}[htbp]
 \centering
\subfigure[]{
   \includegraphics[width=0.8\textwidth]{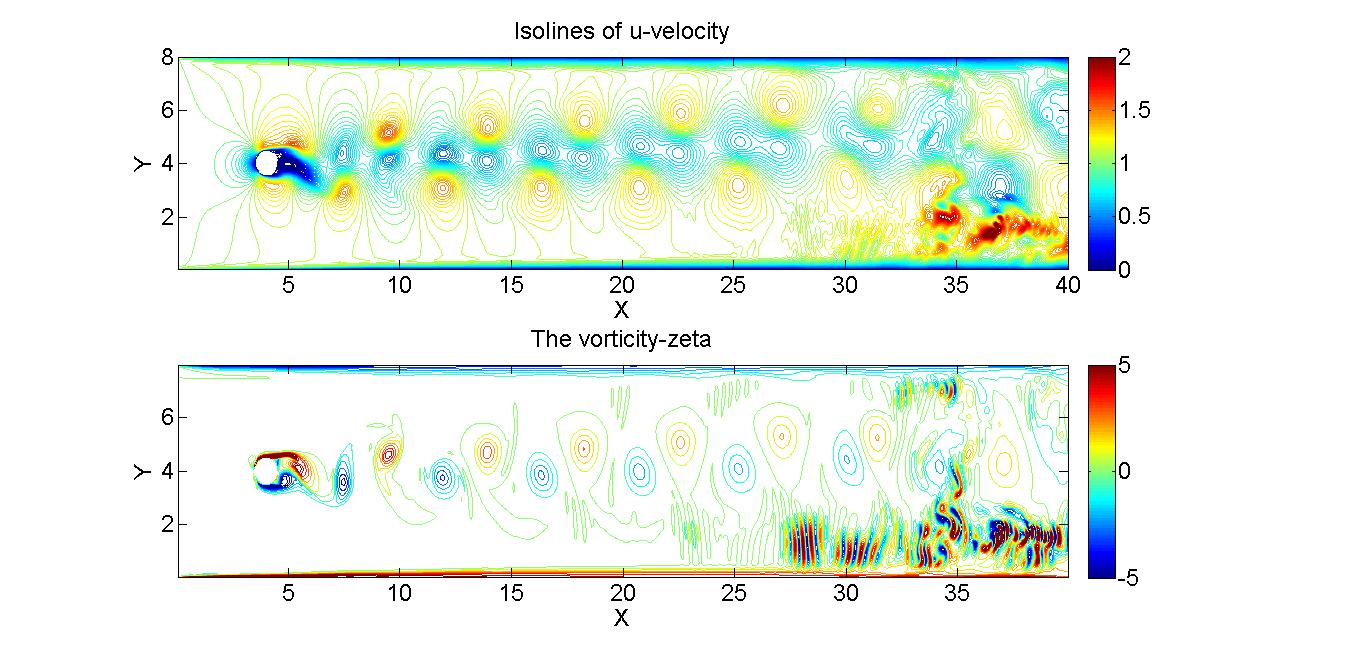} 
  \label{fig-re1500}
 }
 \\
 \subfigure[]{
   \includegraphics[width=0.8\textwidth]{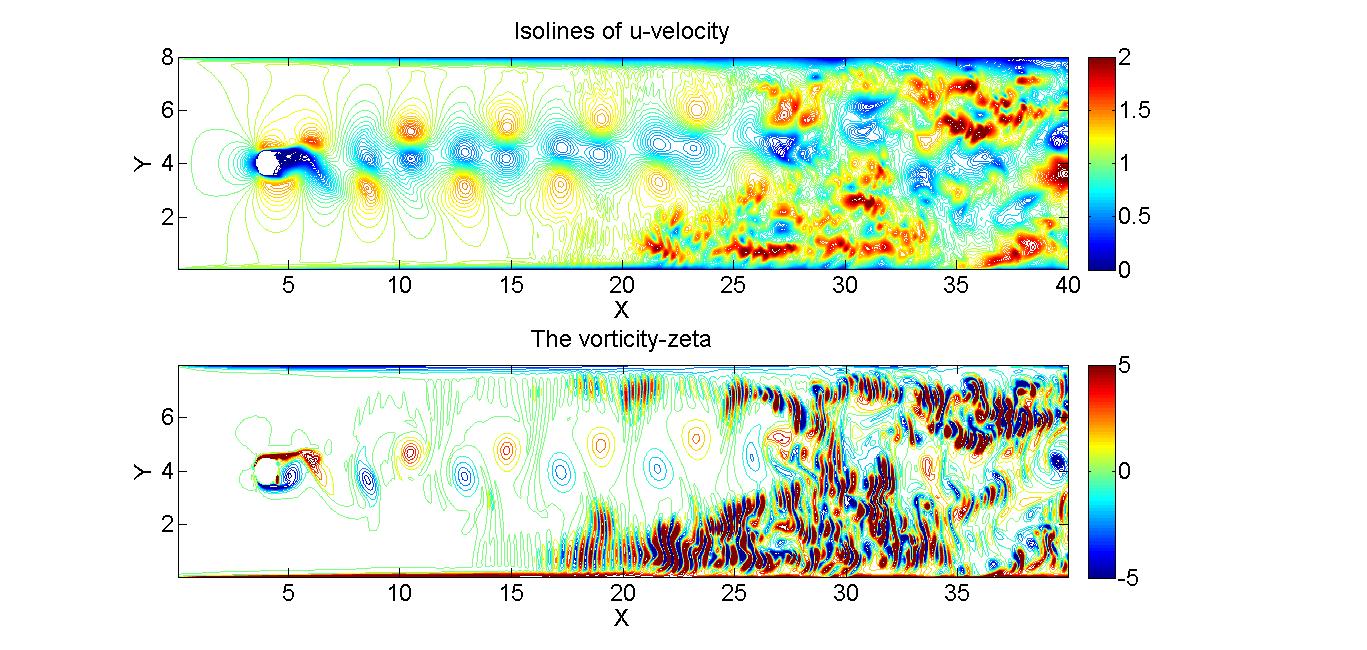} 
  \label{fig-re3000}
 }
 \\
 \subfigure[]{
   \includegraphics[width=0.8\textwidth]{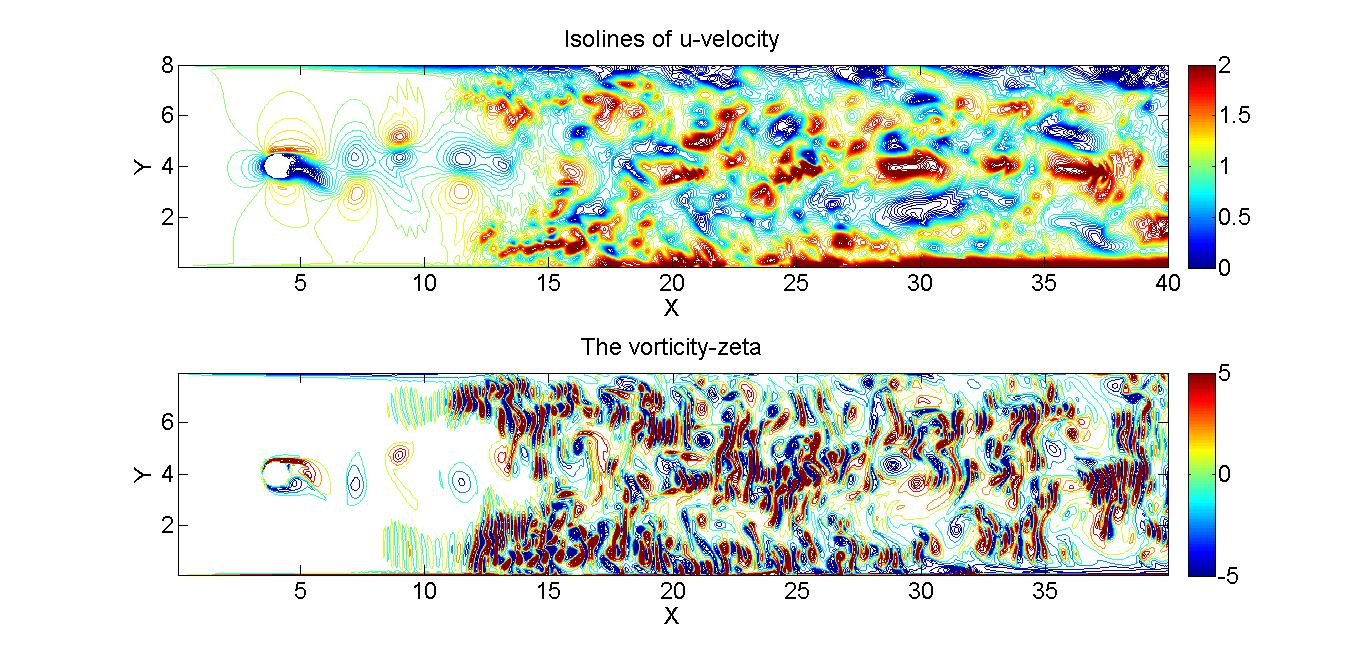} 
   \label{fig-re10000}
 }
 \caption{Contour lines of velocity distribution and vorticity distribution (a) Re=1,500 (b) Re=3,000 (c) Re= 10,000.}
 \label{fig-re-1-2-3}
\end{figure}

Continue to increase $\text{Re}$, the "vortex street" behind the cylinder gradually loses its periodicity in the downstream, and the turbulent core appears in some downstream positions. When the  $\text{Re}$ does not reach large enough, the laminar boundary layer is still near the cylindrical wall before the separation point, and is laminar wake behind the separation point; Further increasing  $\text{Re}$, the turbulent core of the cylindrical wake begins to diffuse upstream, and finally, a complete turbulence is formed. Figure \ref{fig-re-1-2-3} shows the process of turbulence development with the increase of  $\text{Re}$.

The cylinder will experience resistance in fluid flow and is usually evaluated by the drag coefficient. The drag coefficient is calculated by equation .

 $$
C_D=\frac{2F_D}{\rho V^2A}
$$
where $F_D$ is the drag force, $V$ is velocity of the inflow, and $A$ is the projected area of the cylinder perpendicular to the direction of the inflow.

The drag coefficient is mainly dependent on  $\text{Re}$, and is also affected by the position, size, fluid density and viscosity of the cylinder. The drag was calculated using wake integral method, and the calculation results are listed in table 1.

\begin{table}
    \caption{Drag coefficient at different Reynolds numbers}
    \centering
   \begin{tabular}{cccc}
   \toprule
       Re & Experimental data  & Calculation results & RE(\%)\\  
       \midrule
        38 & 1.65 & 1.524 & -7.63  \\ 
        190 & 1.5 & 1.412 & -5.86  \\ 
        1000 & 1.0 & 1.043 & 4.3  \\ 
        $1.00\times10^4$ & 1.05 & 1.102 & 4.95  \\ 
       $1.00\times10^5$ & 1.4 & 1.242 & -11.28  \\ 
        $2.50\times10^5$ & 1.1 & 0.8963 & -18.51  \\ 
        $1.00\times10^6$ & 0.38 & 0.6063 & 59.55  \\ 
        $2.00\times10^6$ & 0.6 & 0.6721 & 12.01  \\ 
        $4.00\times10^6$ & 0.7 & 0.8024 & 14.62 \\ 
        \bottomrule
    \end{tabular}
   \label{tab:tabel1} 
\end{table}

\begin{figure}[htbp]
\centering \includegraphics[width=0.8\textwidth]{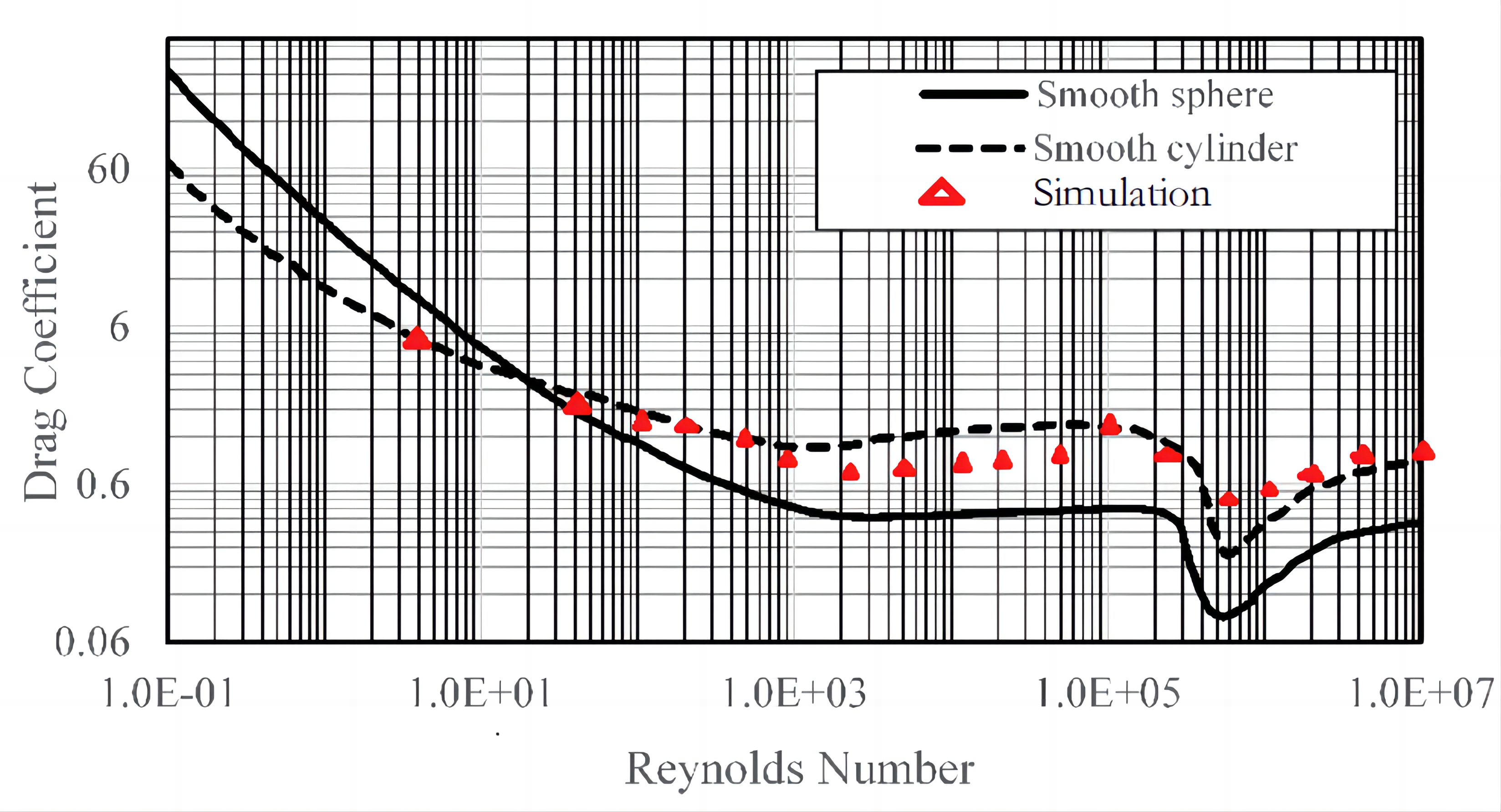} 
\caption{ The variation of drag coefficient with Re.} 
\label{fig9}
\end{figure}

Figure \ref{fig9} shows the variation of drag coefficient with  $\text{Re}$. It can be seen that the simulation results are largely consistent with the experimental data of smooth cylinder and sphere . At low Reynolds number, the drag coefficient decreases linearly with the increase of Re, and the simulation results are in good agreement with the experimental data. The drag coefficient decreases slowly as the  $\text{Re}$ increases from 100 to 2,000.  $\text{Re}$ gradually rises after 3,000, and reaches a peak at about $10^5$. The simulation results also reflect this trend, but are smaller than that of experimental data. When $10^5<\text{Re}<10^6$, the drag coefficient decreases rapidly and then increases. This change is known as a drag crisis, it is attributable to the boundary layer transitions from laminar to turbulent flow, making the separation point move downstream along the surface of the cylinder. The simulation results can also track this trend\cite{yuce2016numerical}.

Comparing the calculated drag coefficient with the experimental data, it is found that the simulation results can racks various changes well on the whole. Especially, the simulations is able to predict the sudden dip that occurred when $10^5<\text{Re}<10^6$ , this indicates that the new turbulence model have ability to predict the boundary layer’s transition from laminar to turbulent. The simulation results were not well close to the experimental data in the drag crisis, this is generally attributed to mesh sizing\cite{heywood1996artificial,holzer2008new,hilton2011influence}. 

\section{Conclusion and Discussion}

The simulation of turbulent flow is a great challenge in CFD. In general, the DNS of turbulent flow is not feasible for the foreseeable future. For RANS and LES, the closure of the governing equations depend on appropriate turbulence model, however, there is no existence of general-purpose turbulence closure models until now. In this paper, we introduced a new model for turbulence simulation. The new turbulence model was based on the characteristics of multi-scale and similarity at different scales in turbulent flow, prior to the computation, the solution domain was decomposed into two-level scales including the large scale and the small scale, there included  mean flow and large scale eddies in large scale and the turbulent fluctuations in small scale. The problem was solved in large scale by the equations of motion and the effect of the turbulent fluctuations on the mean flow was evaluated approximately by using equivalent eddy. Furthermore, the effect of equivalent eddy was decomposed into two parts including convective effect and diffusion effect. The modified Naiver-Stokes equations were established, and ensured the closure of the equations in large scale. Flow around cylinder was numerically investigated and able to obtain flow behavior from low to high Reynolds number.

This method has similar operations as the LES, but there are some differences, in the new  method, an filter function do not need give explicitly; The biggest difference is the modeling of small-scale components between the new method and the LES.

 The application of equivalent vortice has clear physical significance, and the newly introduced scale coefficient ha clear meaning and are easy to determine.
 
This method is universal and able to solve various turbulence problems of incompressible Newtonian fluid flow. Through example verification, this method can establish a unified method for the flow of different Reynolds number and obtain the characteristics of flow from low to high Reynolds number.

This method has less amount of calculation. According to the characteristics of the governing equations, there is no obvious difference in amount of calculation between the laminar and the turbulent flow.

As a new calculation method, the next step also needs to check through a large number of actual flow problems to determine the effectiveness of the method. It is also necessary to further explore the intrinsic nature of the method in the physical sense.

\bibliographystyle{unsrtnat}
\bibliography{references}  






\end{document}